\documentclass{article}
\usepackage{smc2024}
\usepackage{booktabs}
\usepackage{boldline}
\usepackage[caption=false, font=footnotesize]{subfig}
\usepackage{microtype}
\usepackage{paralist}
\usepackage[figure,table]{hypcap}
\usepackage{dblfloatfix} 

\usepackage[whole]{bxcjkjatype}

\usepackage{alphabeta}
\usepackage{arabtex}
\usepackage[LFE,LAE,LGR,T2A,T1]{fontenc}
\usepackage[greek, russian, main=english]{babel}

\usepackage{xcolor}

\def\papertitle{SMUG-Explain: A Framework for \\ Symbolic Music Graph Explanations}

\author[1]{\mbox{\firstname{Emmanouil}\lastname{Karystinaios}}\email{emmanouil.karystinaios@jku.at}}
\author[1,2]{\mbox{\firstname{Francesco}\lastname{Foscarin}}}
\author[1,2]{\mbox{\firstname{Gerhard}\lastname{Widmer}}}

\affil[1]{\department{Institute of Computational Perception}\institution{Johannes Kepler University Linz}\country{Austria}\affiliationtype{University}}
\affil[2]{\department{LIT AI Lab}\institution{Linz Institute of Technology}\country{Austria}\affiliationtype{University}}

\completesetup

\title{\papertitle}
\begin{document}
\capstartfalse
\maketitle
\capstarttrue

\begin{abstract}
    In this work, we present Score MUsic Graph (SMUG)-Explain, a framework for generating and visualizing explanations of graph neural networks applied to arbitrary prediction tasks on musical scores. Our system allows the user to visualize the contribution of input notes (and note features) to the network output, directly in the context of the musical score. We provide an interactive interface based on the music notation engraving library Verovio. We showcase the usage of SMUG-Explain on the task of cadence detection in classical music. 
    All code is available on \url{https://github.com/manoskary/SMUG-Explain}.
\end{abstract}

\section{Introduction}\label{sec:introduction}
In recent years, Graph Neural Networks (GNNs) have emerged as a method for processing musical scores in Music Information Research (MIR) applications, such as cadence detection\cite{karystinaios2022cadence}, expressive performance rendering\cite{jeong2019graph}, optical music recognition~\cite{baro2022musigraph}, music generation~\cite{graphgen}, voice separation\cite{karystinaios2023voice}, and Roman numeral analysis~\cite{karystinaios2023roman}. Like the majority of deep learning-based approaches, GNNs are not intrinsically interpretable, thus making it impossible to inspect the system to reveal potential issues of the system itself and the data it uses or to gain knowledge about the specific task~\cite{molnar2020interpretable}. One could modify the system to make it more interpretable, but this often leads to a reduction in performance. A popular alternative is the so-called \textit{post-hoc} method which aims to explain already trained deep models.

In the field of MIR, multiple explanation techniques have been proposed in recent years~\cite{mishra2017local, mishra2020reliable,Haunschmid2020MML, melchiorre2021ecir,Foscarin2022ConceptBasedTF}, but they all target systems which use matrix-like inputs, such as spectrograms or pianorolls. 
In this paper, instead, we focus on the explainability of MIR systems that use graphs as input and process musical scores. We argue that graph explanations for scores are more musically interpretable since they point to individual note elements and their neighborhood in the score.
We experiment with various post-hoc gradient-based explanation methods for GNNs from the literature~\cite{springenberg2014striving,zeiler2014visualizing,strumbelj2010efficient,simonyan2013deep,faber2021comparing,amara2022graphframex,yuan2022explainability}. 
The generated explanations are quantitatively evaluated by verifying that our explanations satisfy the sufficiency and necessity conditions from a GNN point of view~\cite{amara2022graphframex}. We name our framework Score MUsic Graph (SMUG)-Explain.

Once the explanations are produced, there remains the question of how to present them to the user in an effective way. We display them directly on the musical score, to promote a clear and intuitive relation between the system input, the output and the explanations. Moreover, our use case is more complex than the (more common) explanation of global classifiers (i.e., systems that predict one label for each input excerpt). We want to be able to target systems that predict labels for multiple elements in the input, e.g., for every note or for every time step. This calls for an interactive interface that enables users to focus on specific subsets of predictions.
We start with a web-based graphical rendition of a digitized musical score, based on the engraving library Verovio~\cite{Pugin2014VerovioAL}. The user can interact with individual notes to trigger the visualization of explanations of the target GNN model. Specifically, each note is associated with the subgraph that most contributes to the underlying model's prediction. Furthermore, for each selected note, the user can visualize its feature importance, i.e. the relative importance of input features of the notes involved in the explanation. 

We showcase SMUG-Explain on the cadence detection model of Karystinaios et al.~\cite{karystinaios2022cadence}, by comparing different explanation techniques, and by a qualitative analysis of three music excerpts. GNNs excel at capturing intricate relationships and patterns within graph-structured data. In music, that could be related to voice-leading, harmonic relations, melodic patterns, and other elements that are implicitly modeled by the score graph. 
From a musicological point of view, some parallelism can be found between our explainer and an analyst who highlights the voice leading and most important notes that contribute to the analysis assessment. In particular, some analysis methodologies, such as Schenkerian analysis or GTTM~\cite{lerdahl1996generative}, also create graph structures between notes. In the second part of the paper, we attempt to shed some insight into more musical interpretations of graph explanations. To achieve this, we apply our framework to several music excerpts and comment on the generated explanations. We believe that explanations of deep analytical models might contain some musical pointers that can correlate with expert musical analyses.


\section{Preliminary Concepts}\label{sec:related}
In this section, we present a common task-independent approach (as emerging from recent papers on the topic) to modelling musical scores with GNNs,
then we give some information on explainability techniques for GNNs and their evaluation.

\subsection{GNN-based Approaches on Musical Scores}\label{subsec:graphs}

\begin{figure}[tbp]
    \centering
    \includegraphics[width=0.8\columnwidth]{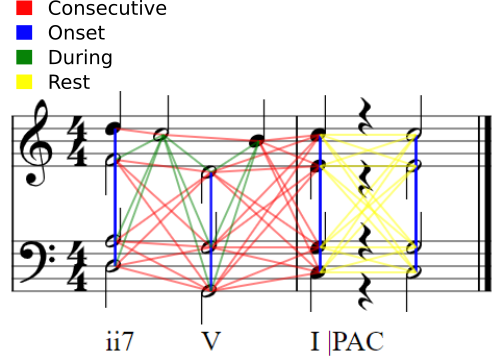}
    \caption{An example of a score graph depicting the different graph edge types in different colours.}
    \label{fig:score_graph}
\end{figure}

The fundamental idea of GNN-based approaches to musical scores is to model a musical score as a graph where notes are the vertices and edges model the temporal relation between the notes.\footnote{More elements could be used as vertices, such as rests, bar lines, and dynamics symbols, but these are not commonly used and we do not consider them in this paper.} The most common approach~\cite{karystinaios2022cadence,karystinaios2023roman,karystinaios2023voice,jeong2019graph} to create a graph from a musical score considers four types of edges (see Figure~\ref{fig:score_graph} for visualization on the score):
\begin{itemize}
    \item \textit{onset edges}: connect notes that share the same onset;
    \item \textit{consecutive edges} (or \textit{next edges}): connect a note $x$ to a note $y$ if the offset of $x$ corresponds to the onset of $y$;
    \item \textit{during edges}: connect a note $x$ to a note $y$ if the onset of $y$ falls within the onset and offset of $x$;
    \item \textit{rest edges} (or \textit{silence edges}): connect the last notes before a rest to the first ones after it.
\end{itemize}
The GNN can treat these four edge types with a single representation~\cite{karystinaios2022cadence}, thus considering a \textit{homogeneous} input graph, or can treat them as different representations per edge type, i.e. as a \textit{heterogeneous} graph.

Adopting the latter approach, a score graph is represented as an attributed heterogeneous graph $G=(V, E, \mathcal{R}, X)$, where $V$ is the set of nodes representing the notes in a score. $E$ is the set of typed edges with elements of the form $(v, \tau, u)$ where, $v, u \in V$ and $r \in \mathcal{R}$ is a relation type. Finally, $X\in \mathcal{R}^{V\times k}$ is a feature matrix such that every node $u$ has its corresponding feature vector $x_u \in X$. Furthermore, we additionally can construct an adjacency matrix $A \in V \times V$.


In our work, we apply the GraphSAGE convolutional block~\cite{hamilton2017inductive}. For a node $v$ the features message passing process per layer $l$ is described as follows:  
\begin{align}\begin{aligned}h_{\mathcal{N}(v)}^{(l+1)} &= \mathrm{aggregate}
\left(\{h_{j}^{l}, \forall j \in \mathcal{N}(v) \}\right)\\h_{v}^{(l+1)} &= \sigma \left(W \cdot \mathrm{concat}
(h_{v}^{l}, h_{\mathcal{N}(v)}^{l+1}) \right)\\h_{v}^{(l+1)} &= \mathrm{norm}(h_{v}^{(l+1)})\end{aligned}\end{align}

Where $W$ is a learnable weight, $\mathcal{N}(v)$ is the set of neighbors of $v$, $\mathrm{aggregate}$ is a permutation invariant aggregation function such as mean, sum, or max, and $\sigma$ is an activation function such as ReLU.


\subsection{Explainability and Graphs}\label{subsec:explainability}

As explained in the introduction, we are interested in \textit{post-hoc} methods, i.e., explainability techniques that work on pre-trained models.
Within the post-hoc realm, the dichotomy of model-aware and model-agnostic explanations emerges. Model-aware methods dissect model parameters for insights, while model-agnostic approaches treat the model as a black box, perturbing inputs to unveil the significance of elements in the output. In this work, we try both categories but find the model-aware techniques to work better for our case.




In terms of explanation evaluation, post-hoc GNN explanations can be measured using the \textit{fidelity metric}~\cite{yuan2022explainability}. The fidelity metric measures the impact of the generated explanatory subgraph on the initial prediction, achieved either by exclusively presenting the subgraph to the model (fidelity-) or by excluding it from the entire graph (fidelity+). 
These fidelity scores capture the ability of an interpretable model to replicate the intrinsic logic of the natural phenomenon or the GNN model.

When it comes to describing post-hoc explanations, we can identify two types of explanations based on their fidelity scores: \textit{necessary} and \textit{sufficient}. A sufficient explanation can be used on its own to reproduce the model's prediction,
and it gets a near-zero negative fidelity score. However, a sufficient explanation is not necessarily unique. On the flip side, a necessary explanation is crucial – removing it from the initial graph changes the model's prediction, like a counterfactual explanation. This type of explanation earns a positive fidelity score close to 1. The ideal situation is found when an explanation is both necessary and sufficient. Amara et al.~\cite{amara2022graphframex} propose to balance the sufficiency and necessity requirements with the \textit{characterization score} which is the weighted harmonic mean of the positive and negative fidelities. Therefore a characterization score close to 1 suggests clear, comprehensive, and informative insights into the model's decisions. 

In this work, we use the characterization score as a means to evaluate different explanation techniques and select the most fitting. It has to be noted that the evaluation of explainability techniques is a particularly complex field, and many approaches have been proposed and then put into question by subsequent research~\cite{hoedt2023constructing,Adebayo2018SanityCF,Yona2021RevisitingSC}. Be that as it may, in this work, it is assumed that the characterization score is a suitable metric to measure the quality of graph explanations.

\section{Our Approach}\label{sec:methodology}
In this section, we detail the cadence detection model that was trained and used to showcase the explanations, then we describe our framework, and finally, we focus on the choice of explanation techniques.

\subsection{Cadence Detection Model}\label{subsec:cad_model}
To showcase our framework, we chose to use a slightly modified version of the cadence detection model introduced by Karystinaios and Widmer~\cite{karystinaios2022cadence}. We extended the model to use as input a heterogeneous score graph as described in Section~\ref{subsec:graphs}. Furthermore, we extended the prediction capabilities of the model from binary (i.e. no-cad or PAC) to multiclass cadence prediction, covering PAC, IAC, and HC. Moreover, we modified the architecture by adding an onset regularization module that sums the latent representations (after the GNN encoder) of all the notes that occur on a distinct onset of the score to every note that shares this onset.
In summary, our cadence detection model consists of a graph convolutional encoder, an onset regularization module, an embedded SMOTE layer for training, and a shallow MLP classifier.

During training, the graph input is passed first through the graph encoder. The obtained node embeddings are then grouped by onset based on the score information, and their representations are averaged together. Following this step, embedded SMOTE is applied to balance the number of cadence classes compared to the notes not having cadence labels in the score. However, when doing inference, the latter synthetic oversampling step is skipped. Finally, the oversampled embeddings are given as input to a shallow $2$-layer MLP classifier that predicts the cadence type.

We trained our model with a joined corpus of cadence annotations from the DCML corpora~\footnote{\url{https://github.com/DCMLab/dcml_corpora}}, the Bach fugues from the well-tempered clavier No. 1~\cite{giraud2015computational}, the annotated Mozart string quartets~\cite{allegraud2019learning}, and the annotated Haydn string quartets~\cite{sears2018simulating}. Our joined corpus makes for $590,149$ individual notes and $17,188$ cadence annotations. We train our model on 90\% of the data and evaluate on 10\% using a random split. Our model reaches a mean F-score of 59\% on the test set. Note that these results cannot be directly compared with \cite{karystinaios2022cadence}, since we use a different (bigger) dataset and perform multiclass prediction.

\subsection{The SMUG-Explain Framework}\label{subsec:smug_model}

\begin{figure*}[btp]
    \centering
    \includegraphics[width=\textwidth]{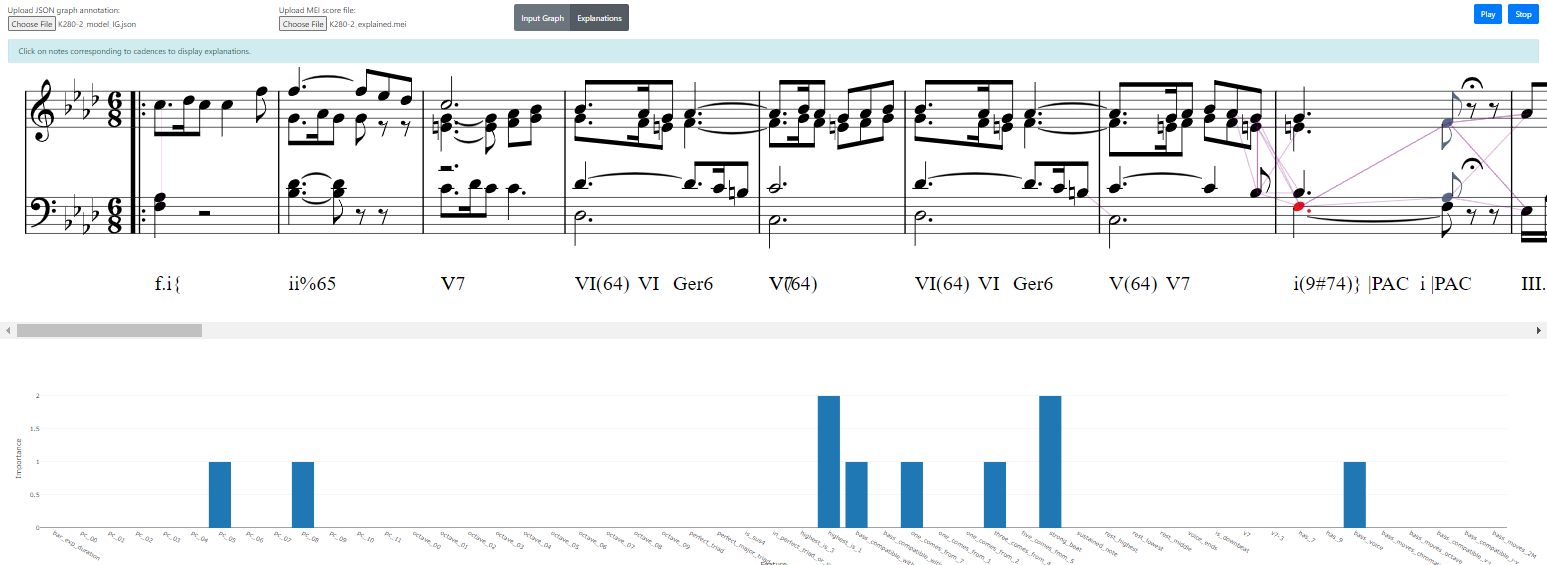}
    \caption{A demonstration of the SMUG-Explain Web interface. In this example, we view the first bars of Mozart's Piano Sonata K280 2$^{nd}$ mvt. It includes a Roman numeral analysis and the cadence label predicted by our model at the top. The purple dashed lines are the produced explanation for the note highlighted in red. Note the vertical connection line in the very first bar, which is also a part of this explanation. At the bottom, we can view the feature importance for the explained note.}
    \label{fig:Mozart_k280.2}
\end{figure*}

Our framework has two main functionalities: generating explanations and making them visually interpretable. 

The first step involves importing musical scores, creating the graph structure, and running the explanation techniques. The score import uses the Python library \textsc{Partitura}~\cite{partitura} which supports a variety of score formats, such as MEI, MusicXML, Kern, or (quantized) MIDI. The graph creation is based on previous work~\cite{karystinaios2023voice, karystinaios2023roman} and outputs a graph in the widely used \textsc{Pytorch Geometric} format~\cite{pyg} to favour reusability and extensibility of our frameworks. 

For the explanation part, we follow a standardized GNN model explanation pipeline: the explainer receives a pre-trained GNN model that performs node-level classification and produces an explanation of the cadence label prediction for a specific note, at two levels: (a) by quantifying the relative importance of the various features of the note; and (b) by identifying an explanation subgraph consisting of the notes and their relations that seem most important for the prediction.
For producing an explanation subgraph, the explainer computes importance masks over all edges. These importance masks reflect an importance score for each one of the edges in the input graph that is used to filter which (most important) edges belong to the explanation subgraph. The importance masks can be soft (i.e. continuous numbers between 0 and 1) or hard (i.e. binary). Furthermore, a maximum number of edges can be imposed per edge type to limit the explanation to stay within a certain graph size. It has been shown that hard masks tend to increase the necessity and sufficiency of explanations~\cite{amara2022graphframex}. Therefore, in our application, we use a hard top-$k$ method for each one of our edge types, where $k$ is set to 10.
Likewise, for producing the feature importance, the explainer produces masks over all nodes and their features and it keeps the $k$ most important by summing the masks along the feature dimension.

The interface of the SMUG-Explain framework is a web-based interface implemented in HTML and Javascript. The core of the interface is the score engraving library \textsc{Verovio}~\cite{Pugin2014VerovioAL} which outputs an SVG representation from a musical score file. We then extend this score representation with the target deep learning model output (i.e., predicted cadences in our case) and with the information produced by the explainers. From the latter, we display in particular the edges between different notes that contribute to the explanation and the feature importance for each note that corresponds to a cadence prediction (see the next section for details). To make this part possible we need a one-to-one mapping between elements in Python and elements in the SVG generated image. Verovio preserves the mapping between note-ids in the musical score file and the SVG image (if they exist), but only the MEI format contains note-ids. Therefore we extended \textsc{Partitura} with an MEI export function, and we support Roman numeral analysis and cadence name exports.

The final step for an effective interface consists of making it interactive. The user can click on single notes, and this will trigger the visualization of the corresponding explanation edges and feature importance. Moreover, the user can switch back and forth between the visualization of the input edges (from the graph presented in Section~\ref{subsec:graphs}) and the explanations, and listen to an audio rendition of the musical score. The interface is shown in Figure~\ref{fig:Mozart_k280.2}.

\subsection{Choice of Explainability Techniques}\label{subsec:evaluation}

To investigate and evaluate the explanations produced by our framework we test several explanation algorithms such as Saliency, Integrated Gradients, Deconvolution, and Guided Backpropagation.
Saliency gauges node and edge importance by weighing each node after calculating the gradient of the output concerning input features~\cite{simonyan2013deep}. Integrated Gradient tackles the saturation issue of gradient-based methods like Saliency by accumulating gradients along the path from a baseline input (zero-vector) to the current input~\cite{sundararajan2017axiomatic}. Deconvolution computes the gradient of the target output but overrides ReLU function gradients, only propagating non-negative gradients~\cite{zeiler2014visualizing}. Guided Backpropagation follows a similar approach, backpropagating only non-negative gradients~\cite{springenberg2014striving}. 

Each of those methods was evaluated on model-level explanations, i.e. the explanation algorithm computes its losses with respect to the model output.
We use the characterization score as defined in \cite{amara2022graphframex} (see Section~\ref{subsec:explainability}). We fix the positive and negative fidelity weights to $0.5$ and apply $k$-top hard masks for edges and nodes.
The results are shown in Table~\ref{tab:explanation_evaluation}.
Our findings suggest that for this task the Integrated Gradients method better captures explanations, in some cases achieving a perfect characterization score of $1$.  

We note that most methods tested on our application are gradient-based. Some perturbation-based methods for generation GNN explanations such as the GNNExplainer~\cite{ying2019gnnexplainer}, Occlusion~\cite{faber2021comparing}, or the GraphMaskExplainer~\cite{schlichtkrull2020interpreting}, might be suitable for our application, however, they are not yet adapted to work with heterogeneous graphs.

As a general remark, we would like to underline that typically graph-dedicated explainers are more biased toward generating compact explanation subgraphs. However, this bias might be less suitable for music where some analysis methods suggest that important elements of a piece might not be directly interconnected.

\begin{table}[h]
\centering
\begin{tabular}{l|cccc}
\toprule
\textbf{Pieces/Model} & \textbf{SAL} & \textbf{GBP} & \textbf{DC} & \textbf{IG} \\
\midrule
WTC-I Fuga 1 & 0.0588 & 0.4706 & 0.2941 & \textbf{1.0} \\
WTC-I Fuga 2 & 0.0 & 0.0435 & 0.0435 & \textbf{0.9130} \\
WTC-I Fuga 5 & 0.0 & 0.0667 & 0.0 & \textbf{0.8667} \\
Mozart K280-2 & 0.0 & 0.3125 & 0.3438 & \textbf{0.9375} \\
Chopin Op.48 & 0.0 & 0.1875 & 0.1250 & \textbf{0.8125} \\
Mozart K331 & 0.0 & 0.0 & 0.0625 & \textbf{1.0} \\
\bottomrule
\end{tabular}
\caption{Characterization score for the model explanations of cadences per piece. The four methods are mentioned in Section~\ref{subsec:evaluation}. SAL stands for Saliency, GBP for Guided Backpropagation, DC for Deconvolution and IG for Integrated Gradients. Highlighted values indicate the highest (best) explanations in terms of characterization score.}
\label{tab:explanation_evaluation}
\end{table}


\section{Qualitative Analysis}\label{sec:examples}

\begin{figure*}[tbp]
    \centering
    \includegraphics[width=\textwidth]{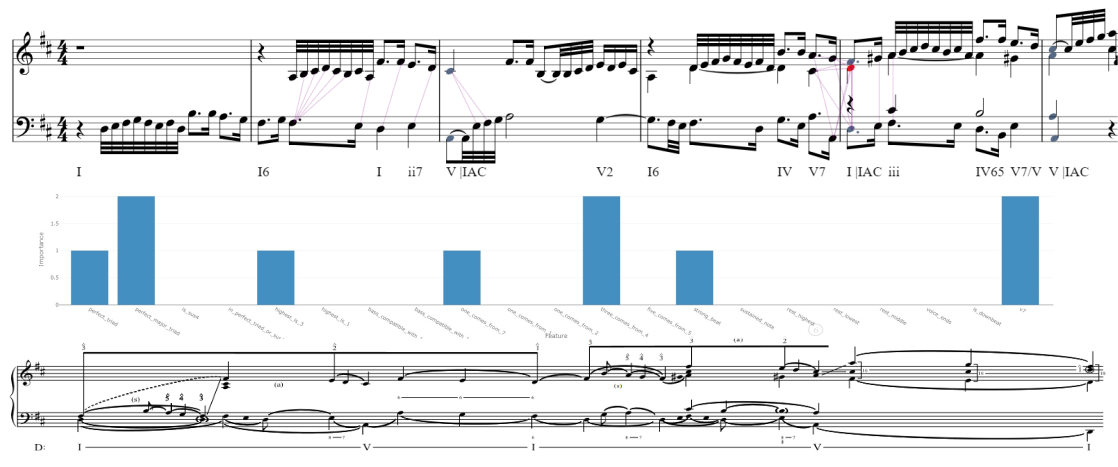}
    \caption{The first bars of the Fuga No.~5 of the Well-Tempered Clavier book No.~1. Top: the score and the explanation of the wrong prediction of the highlighted note in red. Middle: feature importance for the highlighted note. Bottom: a Schenkerian analysis of the segment by~\cite{marlowe2019schenkerian} }
    \label{fig:wtc01f05_explanation}
\end{figure*}

In this section, we perform a qualitative analysis of a diverse set of scores and comment on the graph explanations produced by our system on the task of cadence classification. Apart from the explanation metrics described in Section~\ref{subsec:evaluation} we believe that explanation should be also motivated in musical terms. Therefore, we go into depth about some individual explanations of cadence predictions, targeting both true positive and false positive predictions.\footnote{One could also focus on no-cadence predictions (and false negatives), but we are not considering these in this section, since the ``absence of a specific evidence for a cadence'' may result in less musicologically interesting graph patterns.  
}

We showcase explanations in four classical music excerpts ranging from the Baroque to the Romantic era. Furthermore, we consult expert analyses in terms of harmony, voice leading, and Schenkerian analysis and attempt to interpret the produced explanations. 
Naturally, we assert that the explanations discussed should be necessary and sufficient.
We remind the reader that the generated explanations do not explain the working mechanics of cadences but rather present insights into the model's decision process. In other words, we are not answering the question ``why is there a cadence here?'', but ``why does the model think there is a cadence here?''.

All of the below explanations are included in our publicly available code, including the entire pieces. We invite the reader to explore our interface and test our framework.

\subsection{Mozart Piano Sonata K280 Mov. 2}\label{subsec:mozart}

Our first example (see Figure~\ref{fig:Mozart_k280.2}) focuses on an excerpt from Mozart's Piano Sonata K280 in F major, from the second movement. The harmonic analysis of this segment was provided by Hentschel et al.~\cite{hentschel2021annotated}.
We focus on the perfect authentic cadence arriving on the second beat of measure eight and signaling the end of the first phrase. The harmonic analysis indicates a textbook preparation of the cadence with the German augmented sixth chord on the second beat of measure 6, then the dominant with a tonic in second inversion resolving to a dominant seventh. Finally, the tonic bass arrives on the first beat of measure 8 but with sustained soprano, alto, and tenor voices which themselves resolve on the second beat of the same measure.

We take a closer look at the explanation subgraph of the bass of the tonic (F4) which arrives earlier than the cadential arrival point of the soprano. However, the explanation highlights the descending Urlinie melody containing the $\hat{3}, \hat{2}, \hat{1}$ in the top voice, capturing the first appearance of the $\hat{3}$ on the relative strong beat of the sixth measure (sub-dominant space), whilst considering the overarching bass arpeggiation of the $i-V-i$. Interestingly, the explanation subgraph also contains the first chord of the piece (see faint vertical line in bar 1), indicating that the model considers some information regarding the key of the piece. It's important to highlight that the cadential ground truth obtained from~\cite{hentschel2021annotated} designates notes with the cadence label at the cadential arrival point of the soprano. However, we argue that including the first appearance of the bass note on the strong beat of the same measure is a crucial addition to the cadence annotation ground truth from a musicological standpoint. This addition is accurately captured by our model. Upon analyzing the predictions, we observe that the model correctly identifies the individual notes involved in the PAC in measure 8, namely the $F4$ on the first beat and the soprano, tenor, and alto resolutions on the second beat (remember that our model predicts cadence labels for individual notes, not score onsets).


Furthermore, we deduce from the explanation subgraph that the next notes after the cadence are also crucial for the model's prediction, as they point towards an ending of a phrase and a new rhythmical and harmonic idea further on.


\subsection{Bach WTC Fugue}\label{subsec:bach}

For this example, we explore the predictions and explanations of cadences in the fifth Fuga in D major from J.S.~Bach's Well-tempered Clavier. We believe that the contrapuntal nature of this piece could provide insightful hints about voice leading. Our analysis is complemented by a Schenkerian analysis conducted by Marlowe~\cite{marlowe2019schenkerian}, and we enhance clarity by performing a Roman numeral analysis on the explanatory excerpt (refer to Fig.~\ref{fig:wtc01f05_explanation}).


Our specific focus centers on the explanation of a false positive IAC (Imperfect Authentic Cadence) prediction on the downbeat of the fifth measure. The ground-truth cadence annotations were provided by \cite{giraud2015computational}. The cadence annotations, following the ground truth, were originally provided by Giraud (2015). IAC labels were annotated on the downbeats of measures 3 and 6, both corresponding to the dominant of the D major (i.e. the tonic). Our model accurately predicts these labels but also anticipates an additional IAC label on the downbeat of measure 5, which is not present in the ground truth annotations. Consequently, we investigate what led to this prediction.

We observe that the explanation subgraph for the highlighted red note at the top of Figure~\ref{fig:wtc01f05_explanation} contains, as expected, the edges between the dominant and the subdominant. However, the interesting part of the subgraph takes place in bars 2-3 of the score. The high voice contains the descending melodic line towards the leading tone which then re-appears right before measure 5, and the highlighted red note subsequently resolves it. The lower voice on measure 2 focuses around the E note which would be a fifth from the dominant A. Therefore, the subgraph outlines a movement ii-V-I for the bass.

The middle-ground Schenkerian analysis displayed in the bottom part of Figure~\ref{fig:wtc01f05_explanation} provides some similar observations. We observe an oscillating bass from the tonic to the dominant and over again. This oscillating bass is captured in the cadence annotations by the two cadences on the dominant. It could be argued that the falsely predicted cadence on the tonic strengthens the analysis assumption of the oscillating bass.

In terms of feature importance, we see that characteristics of imperfect cadence are activated such as the existence of a perfect major triad, the highest note being a transposed third interval for the bass, the existence of a leading tone resolving, and the presence of a dominant seventh chord before. Naturally, such characteristics are not exclusively present in the event of cadences but in this case, they seem to influence the model's prediction.

\begin{figure}[tbp]
    \centering
    \includegraphics[width=0.8\columnwidth]{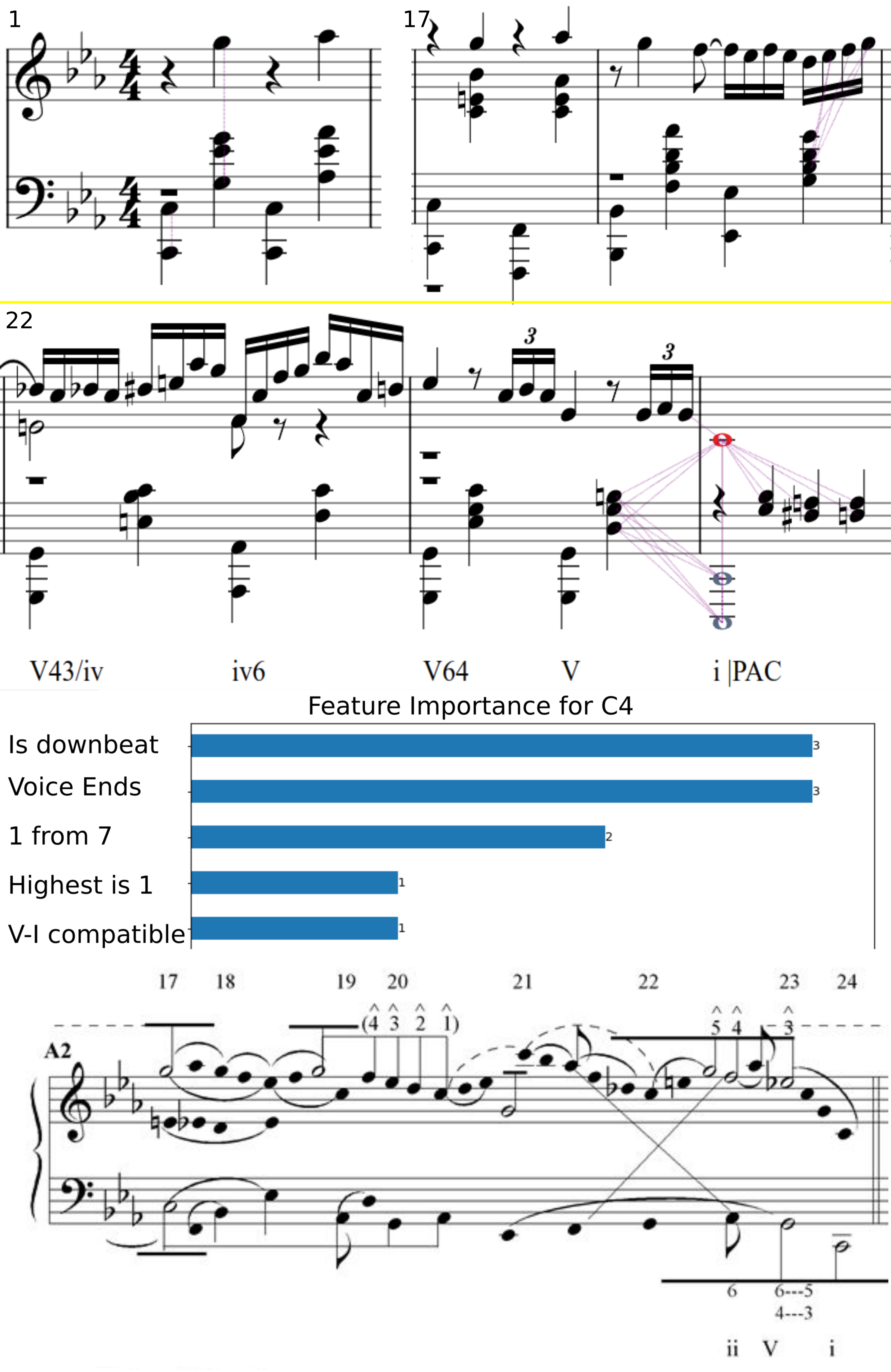}
    \caption{Excerpt of Nocturne Op.~48, no.~1 in C minor by F.~Chopin. Top: excerpts of the explanation for Cadence on measure 24. Bar numbers are notated to the top left of each score segment. Middle: Feature importance for the highlighted C4 note in red. Bottom: Middleground voice leading analysis (from \cite{swinkin2007schenkerian}).}    
    \label{fig:chopin_explaination}
\end{figure}

\subsection{Chopin Nocturne in C minor op. 48}\label{subsec:chopin}


For our last example, we chose a passage with more ambiguous cadential and harmonic elements: bars 22-24 from Chopin's Nocturne op. 48 no 1, in C minor. For this piece, we consult a Schenkerian analysis by Swinkin~\cite{swinkin2007schenkerian}. 
In particular, we look closely at the perfect authentic cadence that arrives in measure 24. Unlike previous examples, this cadence does not have the textbook voice leading and harmonic elements that distinctly define PACs but, nevertheless, carries a cadential character. 

Our model correctly identifies the PAC that arrives on the downbeat of measure 24. To provide support for our analysis and commentary, we perform a Roman numeral analysis on the segment depicted in Figure~\ref{fig:chopin_explaination}. From the model's explanation subgraph, we see that once again the explanation contains the first chord of the piece, further strengthening our assumption that the model is gathering information about the key to inform its predictions. However, the rest of the generated explanation subgraph is mostly compact and local, focusing only on the last notes before the cadence and the ones after it. The chordal content of the segment contains the supported harmony for a perfect authentic cadence (PAC) preparation. But, since the cadence does not follow the typical voice-leading patterns usually involved, it seems that the model does not need to go towards far neighborhoods in the graph to infer the necessary context for its prediction. 

Interestingly, the model includes in the explanation subgraph a part of measure 18. The connected notes in measure 18 correspond to a part of 
the descending Urline that is present in the voice-leading analysis excerpt in the Figure, however, they are not in order. This could be a connection with what one could view as a "sustained" high G in the melody line which slowly descends towards G an octave lower before the cadence. That being said, it is rather ambiguous how the part in measure 18 affects the model's prediction.

\section{Conclusion and Future Work}\label{sec:conclusion}

This paper presented the SMUG-Explain framework for generating and visualizing graph explanations on musical scores. We showcased the framework on a cadence detection model, compared different explanation techniques, and gave some qualitative insights into the explanations. 

Future work will focus on developing new explanation techniques dedicated to musical score graph data, and on testing our results with user-based evaluations. The final objective would be to produce a musicologically trustworthy and user-friendly framework that can support expert analysts to produce more effective musical analyses. Furthermore, we aim to invest efforts into making the system more accessible by releasing an online server-based version of our interface, which can be used to produce predictions for some reference models and tasks and explain them, without the need to run any code locally.




\section{Acknowledgements}
This work is supported by the European Research Council (ERC) under the EU's Horizon 2020 research \& innovation programme, grant agreement No.\ 101019375 (\textit{Whither Music?}), and the Federal State of Upper Austria (LIT AI Lab).

\bibliography{smc2024bib}
	
\end{document}